\documentclass[runningheads]{llncs}
\usepackage[T1]{fontenc}
\usepackage[utf8]{inputenc}
\usepackage{graphicx}
\usepackage{amsmath}
\usepackage{algorithm}
\usepackage{algpseudocode}
\usepackage{biblatex}
\usepackage{amssymb}
\usepackage{float}
\usepackage{array}
\usepackage{booktabs}
\begin{document}

\title{Relative representations for cognitive graphs}

\author{Alex B. Kiefer\inst{1,2} \and Christopher L. Buckley\inst{1,3}}

\authorrunning{Kiefer and Buckley}
\institute{VERSES Research Lab \and Monash University \and Sussex AI Group, Department of Informatics, University of Sussex}

\maketitle
\begin{abstract}
Although the latent spaces learned by distinct neural networks are not generally directly comparable, even when model architecture and training data are held fixed, recent work in machine learning \cite{moschella2023relative} has shown that it is possible to use the similarities and differences among latent space vectors to derive ``relative representations'' with comparable representational power to their ``absolute'' counterparts, and which are nearly identical across models trained on similar data distributions. Apart from their intrinsic interest in revealing the underlying structure of learned latent spaces, relative representations are useful to compare representations across networks as a generic proxy for convergence, and for zero-shot model stitching \cite{moschella2023relative}.\newline

In this work we examine an extension of relative representations to discrete state-space models, using Clone-Structured Cognitive Graphs (CSCGs) \cite{Rikhye864421} for 2D spatial localization and navigation as a test case in which such representations may be of some practical use. Our work shows that the probability vectors computed during message passing can be used to define relative representations on CSCGs, enabling effective communication across agents trained using different random initializations and training sequences, and on only partially similar spaces. In the process, we introduce a technique for zero-shot model stitching that can be applied \emph{post hoc}, without the need for using relative representations during training. This exploratory work is intended as a proof-of-concept for the application of relative representations to the study of cognitive maps in neuroscience and AI.

\keywords{Clone-structured cognitive graphs  \and Relative representations \and Representational similarity}
\end{abstract}

\section{Introduction}

In this short paper we explore the application of relative representations \cite{moschella2023relative} to discrete (graph-structured) models of cognition in the hippocampal-entorhinal system --- specifically, Clone-Structured Cognitive Graphs (CSCGs) \cite{Rikhye864421}. In the first two sections we introduce relative representations and their extension to discrete latent state spaces via continuous messages passed on graphs. We then introduce CSCGs and their use in SLAM (Simultaneous Localization And Mapping). Finally, we report preliminary experimental results using relative representations on CSCGs showing that (a) relative representations can indeed be applied successfully to model the latent space structure of discrete, graph-like representations such as CSCGs, and more generally POMDPs such as those employed in discrete active inference modeling \cite{DACOSTA2020102447, DBLP:journals/corr/abs-2201-03904, SMITH2022102632}; (b) comparison of agents across partially disparate environments reveals important shared latent space structure; and (c) it is possible to use the messages or beliefs (probabilities over states) of one agent to reconstruct the corresponding belief distributions of another via relative representations, without requiring the use of relative representations during training. These examples illustrate an extension of existing representational analysis techniques developed within neuroscience \cite{kriegeskorteetal2008}, which we hope will prove applicable to the study of cognitive maps in biological agents.

\section{Relative representations}

Relative representation \cite{moschella2023relative} is a technique recently introduced in machine learning that allows one to map the intrinsically distinct continuous latent space representations of different models to a common shared representation identical (or nearly so) across the source models, so that latent spaces can be directly compared, even when derived from models with different architectures. The technique is conceptually simple: given anchor points $\mathcal{A} = [\mathbf{x}_1, \mathbf{x}_2, ..., \mathbf{x}_N]$ sampled from a data or observation space and some similarity function $sim$ (e.g. cosine similarity)\footnote{The selection of both suitable anchor points and similarity metrics is discussed at length in \cite{moschella2023relative}. We explain our choices for these hyperparameters in section 5.2 below.}, the relative representation $\mathbf{r}^M_i$ of datapoint $\mathbf{x}_i$ with respect to model $M$ can be defined in terms of $M$'s latent-space embeddings $\mathbf{e}^M_i = f_{enc_M}{(\mathbf{x}_i)}$ as:

\begin{equation}
    \label{rr}
    \mathbf{r}^M_i = [sim(\mathbf{e}^M_i, \mathbf{e}^M_{a_1}), sim(\mathbf{e}^M_i, \mathbf{e}^M_{a_2}), ..., sim(\mathbf{e}^M_i, \mathbf{e}^M_{a_N})]
\end{equation}
where $\mathbf{e}^M_{a_i}$ is the latent representation of anchor $i$ in $M$.

Crucially, the anchor points $\mathcal{A}$ must be matched across models in order for their relative representations to be compatible. ``Matching'' is in the simplest case simply identity, but there are cases in which it is feasible to use pairs of anchors related by a map $g(x) \rightarrow y$ (see below).

In \cite{moschella2023relative} it is shown that the convergence of a model $M_{target}$ during training is well predicted by the average cosine similarity between its relative representations of datapoints and those of an independently validated reference model $M_{ref}$. This is to be expected, given that there is an optimal way of partitioning the data for a given downstream task, and that distinct models trained on the same objective approximate this optimal solution more or less closely, subject to variable factors like random initialization and hyperparameter selection.

While relative representations were recently introduced in machine learning, they take their inspiration in part from prior work on representational similarity analysis (RSA) in neuroscience \cite{kriegeskorteetal2008, DIMSDALEZUCKER2018509}. Indeed, there is a formal equivalence between relative representations and the Representational Dissimilarity Matrices (RDMs) proposed as a common format for representing disparate types of neuroscientific data (including brain imaging modalities as well as simulated neuronal activities in computational models) in \cite{kriegeskorteetal2008}. Specifically, if a similarity rather than dissimilarity metric is employed\footnote{See \cite{kriegeskorteetal2008} fn.2.}, then each row (or, equivalently, column) of the RDM used to characterize a representational space is, simply, a relative representation of the corresponding datapoint. 

Arguably the main contribution of \cite{moschella2023relative} is to exhibit the usefulness of this technique in machine learning, where relative representations may be employed as a novel type of latent space in model architectures. Given a large enough sample of anchor points, relative representations bear sufficient information to play functional roles similar to those of the ``absolute'' representations they model, rather than simply functioning as an analytical tool (e.g. to characterize the structure of latent spaces and facilitate abstract comparisons among systems).

The most obvious practical use of relative representations is in enabling ``latent space communication'': Moschella et al \cite{moschella2023relative} show that the projection of embeddings from distinct models onto the same relative representation enables ``zero-shot model stitching'', in which for example the encoder from one trained model can be spliced to the decoder from another (with the relative representation being the initial layer supplied as input to the decoder). A limitation of this procedure is that it depends on using a relative representation layer during training, precluding its use for establishing communication between ``frozen'' pretrained models. Below, we make use of a parameter-free technique that allows one to map from the relative representation space back to the ``absolute'' representations of the input models with some degree of success.
   
\section{Extending relative representations to discrete state-space models}

Despite the remarkable achievements of continuous state-space models in deep learning systems, discrete state spaces continue to be relevant, both in machine learning applications, where discrete ``world models'' are responsible for state-of-the-art results in model-based reinforcement learning \cite{DBLP:journals/corr/abs-2010-02193}, and in neuroscience, where there is ample evidence for discretized, graph-like representations, for example in the hippocampal-entorhinal system \cite{WHITTINGTON20201249, safron_çatal_verbelen_2021, Rikhye864421} and in models of decision-making processes that leverage POMDPs (Partially Observable Markov Decision Processes) \cite{SMITH2022102632}.

While typical vector similarity metrics such as cosine distance behave in a somewhat degenerate way when applied to many types of discrete representations (e.g., the cosine similarity between two one-hot vectors in the same space is 1 if the vectors are identical and 0 otherwise), they can still be usefully applied in this case (see section 5 below). More generally, the posterior belief distributions inferred over discrete state spaces during simulations in agent-based models may provide suitable anchor points for constructing relative representations.

Concretely, such posterior distributions are often derived using message-passing algorithms, such as belief propagation \cite{10.5555/2876686.2876719} or variational message passing \cite{10.5555/1046920.1088695}. We pursue such a strategy for deriving relative representations of a special kind of hidden Markov model (the Clone-Structured Hidden Markov Model or (if supplemented with actions) Cognitive Graph \cite{Rikhye864421}), in which it is simple to compute forward messages which at each discrete time-step give the probability of the hidden states $z$ conditioned on a sequence of observations $o$ (i.e. $P(z_t|o_{1:t})$). The CSCG/CHMM is particularly interesting both because of its fidelity as a model of hippocampal-entorhinal representations in the brain and because, as in the case of neural networks, distinct agents may learn superficially distinct CSCGs that nonetheless form nearly isomorphic cognitive maps, as shown below.

\section{SLAM using Clone-Structured Cognitive Graphs}

An important strand of research in contemporary machine learning and computational neuroscience has focused on understanding the role of the hippocampus and entorhinal cortex in spatial navigation \cite{stachenfeldetal2017, whitmccafbaker2022, WHITTINGTON20201249, Rikhye864421}, a perspective that may be applicable to navigation in more abstract spaces as well \cite{safron_çatal_verbelen_2021, swaminathan2023schemalearning}. This field of research has given rise to models like the Tolman-Eichenbaum machine \cite{WHITTINGTON20201249} and Clone-Structured Cognitive Graph \cite{george2021clone, Rikhye864421}. We focus on the latter model in the present study, as it is easy to implement on toy test problems and yields a suitable representation for our purposes (an explicit discrete latent space through which messages can be propagated).

The core of the CSCG is a special kind of ``clone-structured'' Hidden Markov Model (CHMM) \cite{Rikhye764456}, in which each of $N$ possible discrete observations are mapped deterministically to only a single ``column'' of hidden states by the likelihood function, i.e. 
$p(o|z) = \begin{cases} 1 & \text{if } z \in C(o) \\ 0 & \text{if } z \notin C(o)  \end{cases}
$, where $C(o)$ is the set of ``clones'' of observation $o$. The clone structure encodes the inductive bias that the same observation may occur within a potentially large but effectively finite number of contexts (i.e. within many distinct sequences of observations), where each ``clone'' functions as a latent representation of $o$ in a distinct context. This allows the model to efficiently encode higher-order sequences \cite{dedieu2019learning} by learning transition dynamics (``lateral'' connections) among the clones. CSCGs supplement this architecture with a set of actions which condition transition dynamics, creating in effect a restricted form of POMDP. 

\begin{figure}[h]
    \centering
    \includegraphics[width=\linewidth]{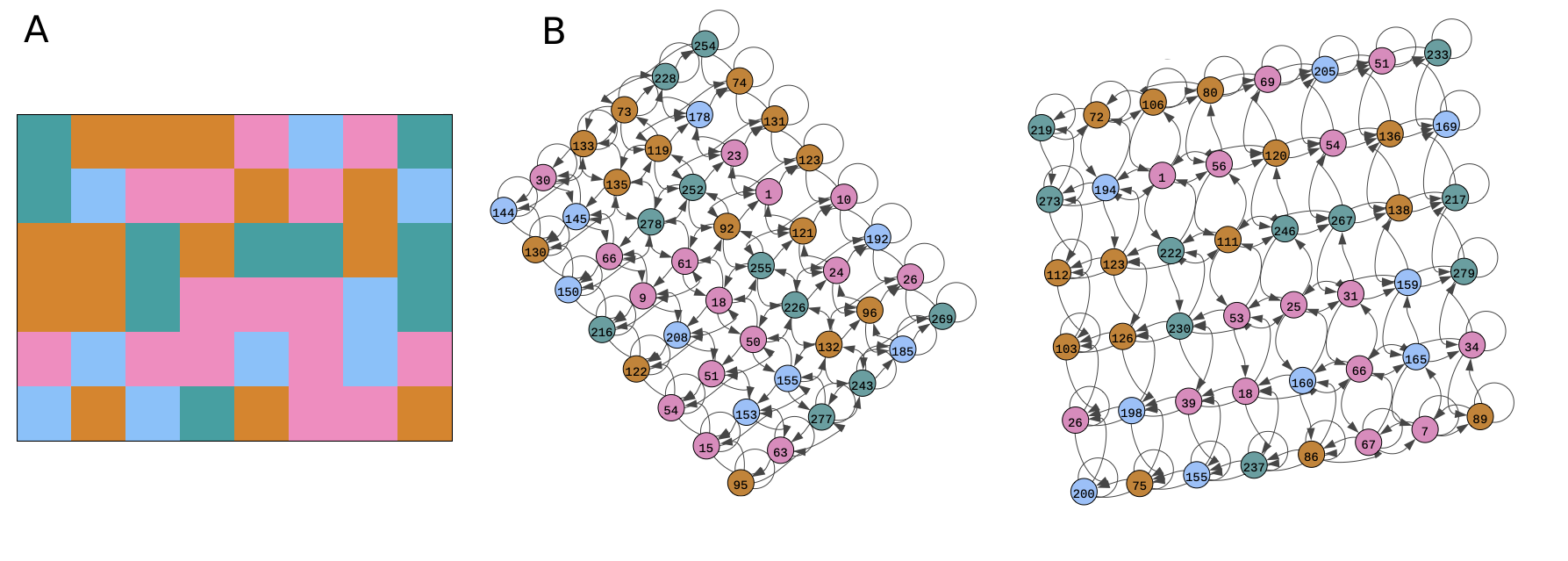}
    \caption{Example of two cognitive graphs (B) learned by CSCG agents via distinct random walks on the same room (A). Following the convention in \cite{Rikhye864421}, colors indicate distinct discrete observations (in the room) or latent ``clones'' corresponding to those observations (in the graphs). Code for training and producing plots is provided in the supplementary materials for \cite{Rikhye864421}. Note that the two graphs are obviously isomorphic upon inspection (the left graph is visually rotated about 50 degrees clockwise relative to the right one, and the node labels differ).}
    \label{learned_cscgs}
    \vspace{-0.5cm}
\end{figure}

The most obvious use of CSCG models (mirroring the function of the hip- pocampal-entorhinal system) is to allow agents capable of moving in a space to perform SLAM (Simultaneous Localization And Mapping) with no prior knowledge of the space's topology. Starting with a random transition matrix, CSCGs trained on random walks in 2D ``rooms'', in which each cell corresponds to an observation, are shown in \cite{Rikhye864421} to be capable of learning action-conditioned transition dynamics among hidden states that exhibit a sparsity structure precisely recapitulating the spatial layout of the room (see Fig. \ref{learned_cscgs}).\footnote{The training used to obtain this result is based on an efficient implementation of the Baum-Welch algorithm for E-M learning, followed by Viterbi training --- please see \cite{Rikhye864421} for details.} 

Given a sequence of observations, an agent can then infer states that correspond to its location in the room, with increasing certainty and accuracy as sequence length increases. Crucially, location is not an input to this model but the agent's representation of location is entirely ``emergent'' from the unsupervised learning of higher-order sequences of observations.

\begin{figure}[h]
    \centering
    \includegraphics[width=\linewidth]{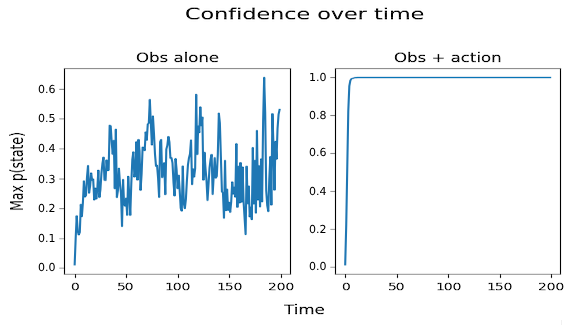}
    \caption{Maximum probability assigned to any hidden state of a CSCG over time (during a random walk). The left panel shows confidence derived from messages inferred from observations alone, and the right panel shows the case of messages inferred from both actions and observations.}
    \label{beliefs}
    \vspace{-0.5cm}
\end{figure}

Building on the codebase provided in \cite{Rikhye864421}, we examined the certainty of agents' inferred beliefs about spatial location during the course of a random walk (see Figure \ref{beliefs}.). Though less than fully confident, such agents are able to reliably infer room location from observation sequences alone after a handful of steps. Conditioning inference as well on the equivalent of ``proprioceptive'' information (i.e., about which actions resulted in the relevant sequence of observations) dramatically increases the certainty of the agents' beliefs. We explored both of these regimes of (un)certainty in our experiments.

\vspace{-0.3cm}
\section{Experiments: Communication across cognitive maps}
\vspace{-0.3cm}

\begin{figure}[H]
    \centering
    \includegraphics[width=\linewidth]{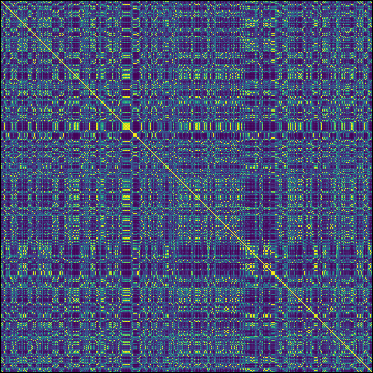}
    \caption{Example representational similarity matrix comparing relative representations of analogous message sequences (i.e. inferred from the same observation sequence) from two distinct models trained on the same environment. This differs from the (dis)similarity matrices typically used in RSA \cite{kriegeskorteetal2008}, as rows and columns in this case represent distinct sets of first-order representations, i.e. cell $(i, j)$ represents the cosine similarity between $\mathbf{r}^A_i$ and $\mathbf{r}^B_j$. Thus the diagonal symmetry illustrates the empirical equivalence of these two sets of relative representations.}
    \label{similarity}
    \vspace{-0.5cm}
\end{figure}

We investigate the extent to which common structure underlying the ``cognitive maps'' learned by distinct CSCG agents can be exploited to enable communication across them. As in the case of neural networks trained on similar data, CSCG agents trained on the same room but with distinct random initializations and observation sequences learn distinct representations that are nonetheless isomorphic at one level of abstraction (i.e. when comparing the structural relationships among their elements, which relative representations make explicit --- cf. Appendix B, Fig. \ref{similarity}). 

We also explore whether partial mappings can be obtained across agents trained on somewhat dissimilar rooms. We used two metrics to evaluate the quality of cross-agent belief mappings: (1) recoverability of the maximum \emph{a posteriori} belief of one agent at a given timestep, given those of another agent following an analogous trajectory; (2) cosine similarity between a given message and its ``reconstruction'' via such a mapping. The main results of these preliminary experiments are reported in Table \ref{results}.

\vspace{-0.3cm}
\subsection{Mapping via permutation}

We first confirmed that CSCG agents trained on distinct random walks of the same room (and with distinct random transition matrix initializations) learn functionally identical cognitive maps if trained to convergence using the procedure specified in \cite{Rikhye864421}. Visualizations of the learned graphs clearly demonstrate topological isomorphism (see references as well as figure \ref{learned_cscgs}B), but in addition we found that the forward messages for a given sequence of observations are identical across agents up to a permutation (i.e., which ``clones'' are used to represent which observation contexts depends on the symmetry breaking induced by different random walks and initializations). It is thus possible to ``translate'' across such cognitive maps in a simple way. First, we obtain message sequences $\mathbf{M}$ and $\mathbf{M'}$ from the first and second CSCGs conditioned on the same observation sequence, and extract messages $\mathbf{m}$ and $\mathbf{m'}$ corresponding to some particular observation $o_t$. We then construct a mapping ${sort\_index}_{\mathbf{m}_{o_t}}(z) \rightarrow {sort\_index}_{\mathbf{m'}_{o_t}}(z')$ from the sort order of entries $z$ in $\mathbf{m}$ to that of entries $z'$ in $\mathbf{m'}$. Using this mapping, we can predict the maximum \emph{a posteriori} beliefs in $\mathbf{M'}$ nearly perfectly given those in $\mathbf{M}$ under ideal conditions (see the ``Permutation (identical)'' condition in Table \ref{results}).\footnote{This procedure does not work if the chosen message represents a state of high uncertainty, e.g. at the first step of a random walk with no informative initial state prior. The mapping also fails for many states since CSCGs, by construction, assign zero probability to all states not within the clone set of a given observation, leading to degeneracy in the mapping. We also found that accuracy of this method degrades rapidly to the extent that the learned map fails to converge to the ground truth room topology.}

\subsection{Mapping via relative representations}

Though it is thus relatively simple to obtain a mapping across cognitive graphs in the ideal case of CSCGs trained to convergence on identical environments, we confirm that relative representations can be used in this setting to obtain comparable results. A message $\mathbf{m'}$ from the second sequence (associated with model B) can be reconstructed from message $\mathbf{m}$ in the first (model A's) by linearly combining model B's embeddings $\mathbf{E}^B_{\mathcal{A}}$ of the anchor points, via a softmax ($\sigma$) function (with temperature $T$) of the relative representation $\mathbf{r}^A_\mathbf{m}$ of $\mathbf{m}$ derived from model A's anchor embeddings:\footnote{In practice, a softmax with a low temperature worked best for reconstruction.}

\begin{equation}
    \label{reconstruction}
    \hat{\mathbf{m'}} = \big(\mathbf{E}^B_\mathcal{A}\big)\sigma\Big[\frac{\mathbf{r}^A_\mathbf{m}}{T}\Big]
\end{equation}

Intuitively, the softmax term scales the contribution of each vector in the set of anchor embeddings to the reconstruction $\hat{\mathbf{m'}}$ in proportion to its relative similarity to the input embedding, so that the reconstruction is a weighted superposition (convex combination) of the anchor points. The reconstruction of a sequence $\mathbf{M'}$ of $m$ $d'$-dimensional messages from an analogous ``source'' sequence $\mathbf{M}$ of $d$-dimensional messages, with the ``batch'' relative representation operation\footnote{If $\mathbf{M} = \mathcal{A}$, this term is a representational similarity matrix in the sense of \cite{kriegeskorteetal2008}.} $\mathbf{R}^A_\mathbf{M} \in \mathbb{R}^{m \times \lvert \mathcal{A} \rvert}$ written out explicitly in terms of the matrix product between $\mathbf{M} \in \mathbb{R}^{{m} \times d}$ and anchor embeddings $\mathbf{E}^A_\mathcal{A} \in \mathbb{R}^{\lvert \mathcal{A} \rvert \times d}$, is then precisely analogous to the self-attention operation in transformers:

\begin{equation}
    \label{batch_reconstruction}
    \hat{\mathbf{M'}} = \sigma\Big[\frac{\mathbf{M}\big[{\mathbf{E}^A_\mathcal{A}}\big]^T}{T}\Big]\mathbf{E}^B_\mathcal{A}
\end{equation}
Here, the source messages $\mathbf{M}$ play the role of the queries $\mathbf{Q}$, model A's anchor embeddings $\mathbf{E}^A_\mathcal{A}$ act as keys $\mathbf{K}$, and model B's anchor embeddings act as values $\mathbf{V}$ in the attention equation which computes output $\mathbf{Z} = \sigma\big[\mathbf{Q}\mathbf{K}^T\big]\mathbf{V}$.\footnote{In the present setting, one might even draw a parallel between the linear projection of transformer inputs to the key, query and value matrices and the linear projection of observations and prior beliefs onto messages via likelihood and transition tensors.} 

Since self-attention may be understood though the lens of its connection to associative memory models \cite{ramsauer2021hopfield, Millidgeetal2022}, this correspondence goes some way toward theoretically justifying our choice of reconstruction method. In particular, following \cite{Millidgeetal2022}, reconstruction via relative representations can be understood as implementing a form of heteroassociative memory in which model A and B's anchor embeddings are, respectively, the memory and projection matrices.

Though empirical performance against a wider range of alternative methods of latent space alignment remains to be assessed, we note a formal connection to regression-based approaches such as \cite{NIPS2002_3a1dd983}, in which a representation $\mathcal{Y}$ of the data is expressed as a mixture of ``guesses'' (linear projections of local embeddings) from $k$ experts, weighted according to the fidelity of each expert's representation of the input data $\mathcal{X}$. This can be expressed as a system of linear equations $\mathcal{Y} = UL$ in which $\mathcal{Y}$, $U$ and $L$ play roles analogous to those of $\hat{\mathbf{M}}$, $\sigma\big[\mathbf{R}^A_\mathbf{M}\big]$ and $\mathbf{E}^B_\mathcal{A}$ above, with the ``repsonsibility'' terms (weights) introducing nonlinearity, as the softmax does in our approach (see Appendix C for further details).

Not surprisingly, the results of our procedure improve with the number of anchors used (see Appendix A, Figure \ref{cosine_sim}). In our experiments, we used $N = 5000$ anchors. We obtained more accurate mappings using this technique when the anchor points were sampled from the trajectory being reconstructed, which raises the probability of an exact match in the anchor set; for generality, all reported results instead sample anchor points (uniformly, without replacement) from distinct random walks. While it would be possible in the present setting to use similarity metrics tailored to probability distributions to create relative representations, we found empirically that replacing cosine similarity with the negative Jensen-Shannon distance slightly adversely affected performance.  

\subsection{Mapping across dissimilar models}

\begin{table}
\begin{center}
\def\arraystretch{1.1}
\setlength\tabcolsep{5.0pt}
\caption{Mapping across distinct CSCG models*}
\label{results}
\begin{tabular}{lc c c} 
 & Max belief recovery & Reconstruction accuracy \\ [0.5ex]
Condition & $\%$ accurate ($\pm$SD) & mean cosine similarity ($\pm$SD) \\ \toprule
 Baseline: AR$^\dag$ (identical) & $0.01 (\pm0.01)$ & $0.07 (\pm0.07)$ \\
 Permutation (identical) & $84.09 (\pm28.9)$ & $0.69 (\pm0.01)$ \\
 Permutation (shifted) & $3.41 (\pm1.48)$ & $0.69 (\pm0.01)$ \\
 Permutation (landmark) & $20.70 (\pm19.14)$ & $0.89 (\pm0.003)$ \\ [3ex]
 \midrule
 RR$^\ddag$ (identical) & $89.44 (\pm1.84)$ & $0.99 (\pm0.003)$ \\
 RR (isomorphic) & $41.0 (\pm3.17)$ & $0.67 (\pm0.02)$ \\
 RR (expansion: large $\rightarrow$ small) & $97.42 (\pm3.24)$ & $0.98 (\pm0.02)$ \\
 RR (expansion: small $\rightarrow$ large) & $47.47 (\pm2.74)$ & $0.59 (\pm0.02)$ \\
 RR (shifted) & $34.81 (\pm3.81)$ & $0.63 (\pm0.03)$ \\
 RR (landmark) & $34.13 (\pm6.47)$ & $0.52 (\pm0.06)$ \\ [1ex] 
 \bottomrule
\end{tabular}
\end{center}
$^\dag$Absolute Representations 
$^\ddag$Relative Representations 
*For each condition, mean results and standard deviation over 100 trials (each run on a distinct random graph) are reported, for the more challenging case of messages conditioned only on observations. For all but the (expansion) conditions, the results of mapping in either direction were closely comparable and we report the mean.
\end{table}

\begin{figure}[h]
    \centering
    \includegraphics[width=\linewidth]{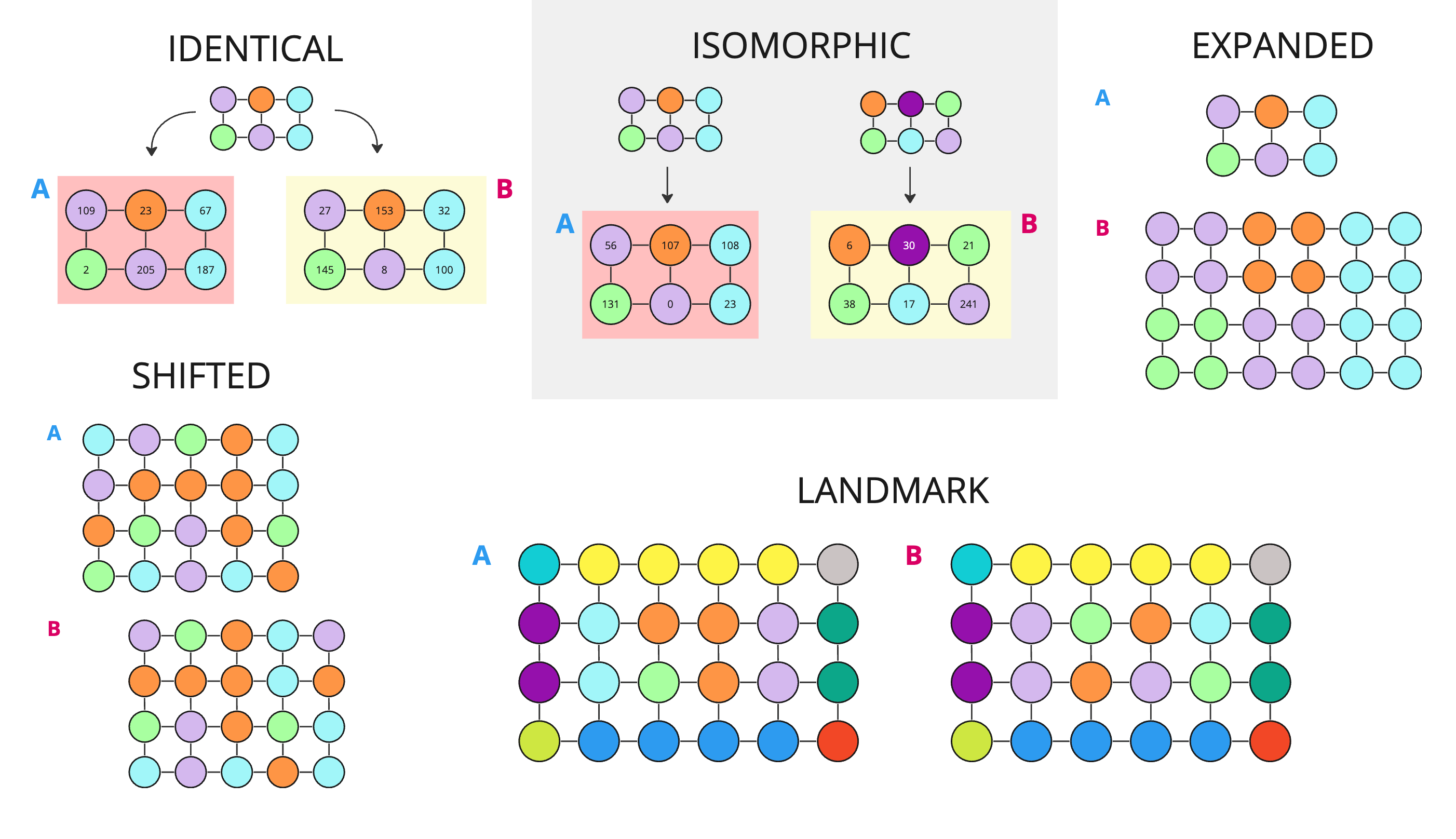}
    \caption{Schematic illustration of experimental conditions. $\mathbf{A}$ and $\mathbf{B}$ indicate distinct rooms on which parallel models were trained, except for the ``IDENTICAL'' condition, where multiple models are trained on a single room. Numbers within nodes illustrate stochastic association of particular hidden state indices with positions in the learned graphs. Graph sizes depicted here do not reflect those used in the experiments.}
    \label{conditions}
    \vspace{-0.5cm}
\end{figure}

As shown in \cite{moschella2023relative}, relative representations can reveal common structure across superficially quite different models --- for example those trained on sentences in distinct natural languages --- via the use of ``parallel'' anchor points, in which the anchors chosen for each model are related by some mapping (e.g. being translations of the same text). In the context of CSCGs, anchors (forward messages) are defined relative to an observation sequence. To sample parallel anchors across agents, we therefore require partially dissimilar rooms in which similar but distinct observation sequences can be generated. 

We used four experimental manipulations to generate pairs of partially dissimilar rooms (see Figure \ref{conditions}), which we now outline along with a brief discussion of our results on each.

\subsubsection{Isomorphism}
Any randomly generated grid or ``room'' of a given fixed size will (if CSCG training converges) yield a cognitive map with the same topology. It should thus be possible to generate parallel sequences of (action, observation) pairs --- and thus parallel anchor points for defining relative representations --- across two such random rooms, even if each contains a distinct set of possible observations or a different number of clones, either of which would preclude the use of a simple permutation-based mapping. 

The relationships among observations will differ across such rooms, however, which matters under conditions of uncertainty, since every clone of a given observation will be partially activated when that observation is received, leading to different conditional belief distributions. This effect should be mitigated or eliminated entirely when beliefs are more or less certain, in which case ``lateral'' connections (transition dynamics) select just one among the possible clones corresponding to each observation. Indeed, we found that it is possible to obtain near-perfect reconstruction accuracy across models trained on random rooms with distinct observation sets, provided that messages are conditioned on both actions and observations; whereas we only obtained a $<50\%$ success rate in this scenario when conditioning on observations alone. 

\subsubsection{Expansion}
In this set of experiments, we generated ``expanded'' versions of smaller rooms and corresponding ``stretched'' trajectories (paired observation and action sequences) using Kroenecker products, so that each location in the smaller room is expanded into a $2 \times 2$ block in the larger room, and each step in the smaller room corresponds to two steps in the larger one. We can then define parallel anchors across agents trained on such a pair of rooms, by taking (a) all messages in the smaller room, and (b) every other message in the larger one. In this condition, the large $\rightarrow$ small mapping can be performed much more accurately than the opposite one, since each anchor point in the smaller (``downsampled'') room corresponds to four potential locations in the larger. Superior results on the (large $\rightarrow$ small) condition VS our experiments on identical rooms may be explained by the fact that the ``small'' room containts fewer candidate locations than the room used in the ``Identical'' condition.

\subsubsection{Shifting}
In a third set of experiments, we generated rooms by taking overlapping vertical slices of a wider room, such that identical sequences were observed while traversing the rooms, but within different wider contexts. In this case only the messages corresponding to overlapping locations were used as anchor points, but tests were performed on random walks across the entire room. Under conditions of certainty, mapping across these two rooms can be solved near-perfectly by using all messages as candidate anchor points, since the rooms are isomorphic. Without access to ground-truth actions, it was possible to recover the beliefs of one agent given the other's only $\sim 35\%$ of the time, even if anchors were sampled from all locations. We hypothesize that this problem is more challenging than the ``Isomorphic'' condition because similar patterns of observations (and thus similar messages) correspond to distinct locations across the two rooms, which should have the effect of biasing reconstructions toward the wrong locations.  

\subsubsection{Landmarks}
Finally, partially following the experiments in \cite{Rikhye864421} on largely featureless rooms with unique observations corresponding to unique locations (e.g. corners and walls), we define pairs of rooms with the same (unique) observations assigned to elements of the perimeter, filled by otherwise randomly generated observations that differed across rooms. Using only the common ``landmark'' locations as anchors, it was still possible to use relative representations to recover an agent's location from messages in a parallel trajectory in the other room with some success.

\subsubsection{Summary}

The results reported in Table \ref{results} were obtained under conditions of significant uncertainty, in which messages were conditioned only on observations, without knowledge of the action that produced those observations. In this challenging setting, relative representations still enabled recovery (well above chance in all experimental conditions, and in some cases quite accurate) of one agent's maximum \emph{a posteriori} belief about its location from those of the other agent, averaged across messages in a test sequence.\footnote{It is worth noting that this is essentially a one-of-N classification task, with effective values of N around 48 in most cases. This is because (following \cite{Rikhye864421}) most experiments were performed on $6 \times 8$ rooms, and there is one ``active'' clone corresponding to each location in a converged CSCG.}

In all settings, it was possible to obtain highly accurate mappings ($>99\%$ correct in most cases) by conditioning messages on actions as well as observations. This yields belief vectors sharply peaked at the hidden state corresponding to an agent's location on the map. In this regime, the reconstruction procedure acts essentially as a lookup table, as a given message $\mathbf{m}$ resembles a one-hot vector and this sparsity structure is reflected in the relative representation (which is $\sim0$ everywhere except for dimensions corresponding to anchor points nearly identical to $\mathbf{m}$). The softmax weighting then simply ``selects'' the corresponding anchor in model B's anchor set.\footnote{There is a variation on this in which multiple matches exist in the anchor set, but the result is the same as we then combine $n$ identical anchor points.} Conditioning messages on probabilistic knowledge of actions (perhaps the most realistic scenario) can be expected to greatly improve accuracy relative to the observation-only condition, and is an interesting subject for a follow-up study.

\section{Discussion}

The ``messages'' used to define relative representations in the present work can be interpreted as probability distributions, but they can also be interpreted more agnostically as, simply, neuronal activity vectors. Recent work in systems neuroscience \cite{dabagiaetal2023} has shown that it is possible to recover common abstract latent spaces from real neuronal activity profiles. As noted above, relative representations were anticipated in neuroscience by RSA, which in effect treats the neuronal responses, or computational model states, associated with certain fixed stimuli as anchor points. This technique complements others such as the analysis of attractor dynamics \cite{doi:10.1126/science.1108905} as a tool to investigate properties of latent spaces in brains, and has been shown to be capable of revealing common latent representational structure across not only individuals, but linguistic communities \cite{Zinszer2016SemanticSA} and even species \cite{Kriegeskorte2008MatchingCO, Haxby2014DecodingNR}. Consistent with the aims of \cite{moschella2023relative} and \cite{kriegeskorteetal2008}, this paradigm might ultimately provide fascinating future directions for brain imaging studies of navigational systems in the hippocampal-entorhinal system and elsewhere. 

Relative representations generalize this paradigm to ``parallel anchors'', and also demonstrate the utility of high-dimensional representational similarity vectors as latent representations in their own right, which can, as demonstrated above, be used to establish zero-shot communication between distinct models. 

While the conditions we constructed in our toy experiments are artificial, they have analogues in more realistic scenarios. It is plausible that animals navigating structurally homeomorphic but superficially distinct environments, for example, should learn similar cognitive maps at some level of abstraction. Something analogous to the ``expansion'' setting may occur across two organisms that explore the same space but (for example due to different sizes or speeds of traversal, and thus sample rates) coarse-grain it differently. The idea of landmark-based navigation is central to the SLAM paradigm generally, and the stability of landmarks across otherwise different spaces may provide a model for the ability to navigate despite changes to the same environment over time. Finally, while experiments on partially overlapping rooms seem somewhat contrived if applied naively to spatial navigation scenarios, they may be quite relevant to models of SLAM in abstract spaces \cite{safron_çatal_verbelen_2021}, such as during language acquisition, where different speakers of the same language may be exposed to partially disjoint sets of stimuli, corresponding to different dialects (or in the limit, idiolects).

Crucially,  the common reference frame provided by these techniques might allow for the analysis of \emph{shared} representations, which (when derived from well-functioning systems) should embody an ideal structure that individual cognitive systems in some sense aim to approximate, allowing for comparison of individual brain-bound models against a shared, abstract ground truth. Such an abstracted ``ideal'' latent space could be used to measure error or misrepresentation \cite{Kiefer2019-KIERIT}, or to assess progress in developmental contexts.

\section{Conclusion}

In this work we have considered a toy example of the application of relative representations to graph-structured cognitive maps. The results reported here are intended mainly to illustrate concrete directions for the exploration of the latent structure of cognitive maps using relative representations, and as a proof-of-principle that the technique can be applied to the case of inferred posterior distributions over discrete latent spaces. We have also introduced a technique for reconstructing ``absolute'' representations from their relative counterparts without learning.

In addition to further investigating hyperparameter settings (such as choice of similarity function) to optimize performance in practical applications, future work might explore the application of relative representations to more complex models with discrete latent states, such as the discrete ``world models'' used in cutting-edge model-based reinforcement learning \cite{DBLP:journals/corr/abs-2010-02193}, or to enable belief sharing and cooperation in multi-agent active inference scenarios. Given the connection to neural self-attention described above, which has also been noted in the context of the Tolman-Eichenbaum Machine \cite{DBLP:journals/corr/abs-2112-04035}, it would also be intriguing to explore models in which such a translation process occurs within agents themselves, as a means of transferring knowledge across local cognitive structures.

\section*{Acknowledgements}
Alex Kiefer is supported by VERSES Research. CLB is supported by BBRSC grant number BB/P022197/1 and by Joint Research with the National Institutes of Natural Sciences (NINS), Japan, program No. 0111200.

\section*{Code Availability}

The CSCG implementation is based almost entirely on the codebase provided in \cite{Rikhye864421}. Code for reproducing our experiments and analysis can be found at: 
\\\text{https://github.com/exilefaker/cscg-rr}

\printbibliography

@misc{moschella2023relative,
      title={Relative representations enable zero-shot latent space communication}, 
      author={Luca Moschella and Valentino Maiorca and Marco Fumero and Antonio Norelli and Francesco Locatello and Emanuele Rodolà},
      year={2023},
      eprint={2209.15430},
      archivePrefix={arXiv},
      primaryClass={cs.LG}
}

@article {Rikhye864421,
	author = {Rajeev V. Rikhye and Nishad Gothoskar and J. Swaroop Guntupalli and Antoine Dedieu and Miguel L{\'a}zaro-Gredilla and Dileep George},
	title = {Learning cognitive maps as structured graphs for vicarious evaluation},
	elocation-id = {864421},
	year = {2020},
	doi = {10.1101/864421},
	publisher = {Cold Spring Harbor Laboratory},
	URL = {https://www.biorxiv.org/content/early/2020/06/24/864421},
	eprint = {https://www.biorxiv.org/content/early/2020/06/24/864421.full.pdf},
	journal = {bioRxiv}
}

@article{DBLP:journals/corr/abs-2010-02193,
  author       = {Danijar Hafner and
                  Timothy P. Lillicrap and
                  Mohammad Norouzi and
                  Jimmy Ba},
  title        = {Mastering Atari with Discrete World Models},
  journal      = {CoRR},
  volume       = {abs/2010.02193},
  year         = {2020},
  url          = {https://arxiv.org/abs/2010.02193},
  eprinttype    = {arXiv},
  eprint       = {2010.02193},
  bibsource    = {dblp computer science bibliography, https://dblp.org}
}

@article{SMITH2022102632,
title = {A step-by-step tutorial on active inference and its application to empirical data},
journal = {Journal of Mathematical Psychology},
volume = {107},
pages = {102632},
year = {2022},
issn = {0022-2496},
doi = {https://doi.org/10.1016/j.jmp.2021.102632},
url = {https://www.sciencedirect.com/science/article/pii/S0022249621000973},
author = {Ryan Smith and Karl J. Friston and Christopher J. Whyte}
}

@misc{safron_çatal_verbelen_2021,
 title={Generalized Simultaneous Localization and Mapping (G-SLAM) as unification framework for natural and artificial intelligences: towards reverse engineering the hippocampal/entorhinal system and principles of high-level cognition},
 url={psyarxiv.com/tdw82},
 DOI={10.31234/osf.io/tdw82},
 publisher={PsyArXiv},
 author={Safron, Adam and Çatal, Ozan and Verbelen, Tim},
 year={2021},
 month={Oct}
}

@article{dabagiaetal2023,
title = {Aligning latent representations of neural activity},
author={Max Dabagia and Konrad P. Kording and Eva L. Dyer},
journal = {Nature Biomedical Engineering},
volume = {7},
year={2023},
month={April},
pages={337-343},
DOI={https://doi.org/10.1038/s41551-022-00962-7},
}

@article{DBLP:journals/corr/abs-2201-03904,
  author       = {Conor Heins and
                  Beren Millidge and
                  Daphne Demekas and
                  Brennan Klein and
                  Karl J. Friston and
                  Iain D. Couzin and
                  Alexander Tschantz},
  title        = {pymdp: {A} Python library for active inference in discrete state spaces},
  journal      = {CoRR},
  volume       = {abs/2201.03904},
  year         = {2022},
  url          = {https://arxiv.org/abs/2201.03904},
  eprinttype    = {arXiv},
  eprint       = {2201.03904},
  bibsource    = {dblp computer science bibliography, https://dblp.org}
}

@article{DACOSTA2020102447,
title = {Active inference on discrete state-spaces: A synthesis},
journal = {Journal of Mathematical Psychology},
volume = {99},
pages = {102447},
year = {2020},
issn = {0022-2496},
doi = {https://doi.org/10.1016/j.jmp.2020.102447},
url = {https://www.sciencedirect.com/science/article/pii/S0022249620300857},
author = {Lancelot {Da Costa} and Thomas Parr and Noor Sajid and Sebastijan Veselic and Victorita Neacsu and Karl Friston}
}

@article{Haxby2014DecodingNR,
  title={Decoding neural representational spaces using multivariate pattern analysis.},
  author={James V. Haxby and Andrew C. Connolly and J. Swaroop Guntupalli},
  journal={Annual review of neuroscience},
  year={2014},
  volume={37},
  pages={
          435-56
        },
  url={https://api.semanticscholar.org/CorpusID:6794418}
}

@inproceedings{NIPS2002_3a1dd983,
 author = {Teh, Yee and Roweis, Sam},
 booktitle = {Advances in Neural Information Processing Systems},
 editor = {S. Becker and S. Thrun and K. Obermayer},
 pages = {},
 publisher = {MIT Press},
 title = {Automatic Alignment of Local Representations},
 url = {https://proceedings.neurips.cc/paper_files/paper/2002/file/3a1dd98341fafc1dfe9bcf36360e6b84-Paper.pdf},
 volume = {15},
 year = {2002}
}

@misc{ramsauer2021hopfield,
      title={Hopfield Networks is All You Need}, 
      author={Hubert Ramsauer and Bernhard Schäfl and Johannes Lehner and Philipp Seidl and Michael Widrich and Thomas Adler and Lukas Gruber and Markus Holzleitner and Milena Pavlović and Geir Kjetil Sandve and Victor Greiff and David Kreil and Michael Kopp and Günter Klambauer and Johannes Brandstetter and Sepp Hochreiter},
      year={2021},
      eprint={2008.02217},
      archivePrefix={arXiv},
      primaryClass={cs.NE}
}

@inproceedings{Millidgeetal2022,
title = {Universal Hopfield Networks: A General Framework for Single-Shot Associative Memory Models},
author = {Beren Millidge and Tommaso Salvatori and Yuhang Song and Thomas Lukasiewicz and Rafal Bogacz},
year = {2022},
month = {July},
pages = {15561-15583},
volume = {162},
location={Baltimore, Maryland, USA},
booktitle={Proceedings of the 39th International Conference on Machine Learning}
}

@misc{swaminathan2023schemalearning,
      title={Schema-learning and rebinding as mechanisms of in-context learning and emergence}, 
      author={Sivaramakrishnan Swaminathan and Antoine Dedieu and Rajkumar Vasudeva Raju and Murray Shanahan and Miguel Lazaro-Gredilla and Dileep George},
      year={2023},
      eprint={2307.01201},
      archivePrefix={arXiv},
      primaryClass={cs.CL}
}

@article{Kriegeskorte2008MatchingCO,
  title={Matching Categorical Object Representations in Inferior Temporal Cortex of Man and Monkey},
  author={Nikolaus Kriegeskorte and Marieke Mur and Douglas A. Ruff and Roozbeh Kiani and Jerzy Bodurka and Hossein Esteky and Keiji Tanaka and Peter A. Bandettini},
  journal={Neuron},
  year={2008},
  volume={60},
  pages={1126-1141},
  url={https://api.semanticscholar.org/CorpusID:313180}
}

@article{Zinszer2016SemanticSA,
  title={Semantic Structural Alignment of Neural Representational Spaces Enables Translation between English and Chinese Words},
  author={Benjamin D. Zinszer and Andrew James Anderson and Olivia Kang and Thalia Wheatley and Rajeev D. S. Raizada},
  journal={Journal of Cognitive Neuroscience},
  year={2016},
  volume={28},
  pages={1749-1759},
  url={https://api.semanticscholar.org/CorpusID:577366}
}

@article{DBLP:journals/corr/abs-2112-04035,
  author       = {James C. R. Whittington and
                  Joseph Warren and
                  Timothy Edward John Behrens},
  title        = {Relating transformers to models and neural representations of the
                  hippocampal formation},
  journal      = {CoRR},
  volume       = {abs/2112.04035},
  year         = {2021},
  url          = {https://arxiv.org/abs/2112.04035},
  eprinttype    = {arXiv},
  eprint       = {2112.04035},
  bibsource    = {dblp computer science bibliography, https://dblp.org}
}

@article{WHITTINGTON20201249,
title = {The Tolman-Eichenbaum Machine: Unifying Space and Relational Memory through Generalization in the Hippocampal Formation},
journal = {Cell},
volume = {183},
number = {5},
pages = {1249-1263.e23},
year = {2020},
issn = {0092-8674},
doi = {https://doi.org/10.1016/j.cell.2020.10.024},
url = {https://www.sciencedirect.com/science/article/pii/S009286742031388X},
author = {James C.R. Whittington and Timothy H. Muller and Shirley Mark and Guifen Chen and Caswell Barry and Neil Burgess and Timothy E.J. Behrens}
}

@article{10.5555/1046920.1088695,
author = {Winn, John and Bishop, Christopher M.},
title = {Variational Message Passing},
year = {2005},
issue_date = {12/1/2005},
publisher = {JMLR.org},
volume = {6},
issn = {1532-4435},
journal = {J. Mach. Learn. Res.},
month = {dec},
pages = {661–694},
numpages = {34}
}

@inproceedings{10.5555/2876686.2876719,
author = {Pearl, Judea},
title = {Reverend Bayes on Inference Engines: A Distributed Hierarchical Approach},
year = {1982},
publisher = {AAAI Press},
booktitle = {Proceedings of the Second AAAI Conference on Artificial Intelligence},
pages = {133–136},
numpages = {4},
location = {Pittsburgh, Pennsylvania},
series = {AAAI'82}
}

@article{whitmccafbaker2022,
author = {Whittington, James and McCaffary, David and Bakermans, Jacob and Behrens, Timothy},
year = {2022},
month = {09},
pages = {1-16},
title = {How to build a cognitive map},
volume = {25},
journal = {Nature Neuroscience},
doi = {10.1038/s41593-022-01153-y}
}

@article{stachenfeldetal2017,
author = {Stachenfeld, Kimberly and Botvinick, Matthew and Gershman, Samuel},
year = {2017},
month = {07},
pages = {},
title = {The hippocampus as a predictive map},
doi = {10.1101/097170}
}

@article{george2021clone,
  title={Clone-structured graph representations enable flexible learning and vicarious evaluation of cognitive maps},
  author={George, Dileep and Rikhye, Rajeev V and Gothoskar, Nishad and Guntupalli, J Swaroop and Dedieu, Antoine and L{\'a}zaro-Gredilla, Miguel},
  journal={Nature communications},
  volume={12},
  number={1},
  pages={2392},
  year={2021},
  publisher={Nature Publishing Group UK London}
}

@misc{dedieu2019learning,
      title={Learning higher-order sequential structure with cloned HMMs}, 
      author={Antoine Dedieu and Nishad Gothoskar and Scott Swingle and Wolfgang Lehrach and Miguel Lázaro-Gredilla and Dileep George},
      year={2019},
      eprint={1905.00507},
      archivePrefix={arXiv},
      primaryClass={stat.ML}
}

@article {Rikhye764456,
	author = {Rajeev V. Rikhye and J. Swaroop Guntupalli and Nishad Gothoskar and Miguel L{\'a}zaro-Gredilla and Dileep George},
	title = {Memorize-Generalize: An online algorithm for learning higher-order sequential structure with cloned Hidden Markov Models},
	elocation-id = {764456},
	year = {2019},
	doi = {10.1101/764456},
	publisher = {Cold Spring Harbor Laboratory},
	URL = {https://www.biorxiv.org/content/early/2019/09/10/764456},
	eprint = {https://www.biorxiv.org/content/early/2019/09/10/764456.full.pdf},
	journal = {bioRxiv}
}

@article{kriegeskorteetal2008,
author = {Kriegeskorte, Nikolaus and Mur, Marieke and Bandettini, Peter},
year = {2008},
month = {02},
pages = {4},
title = {Representational Similarity Analysis – Connecting the Branches of Systems Neuroscience},
volume = {2},
journal = {Frontiers in systems neuroscience},
doi = {10.3389/neuro.06.004.2008}
}

@incollection{DIMSDALEZUCKER2018509,
title = {Chapter 27 - Representational Similarity Analyses: A Practical Guide for Functional MRI Applications},
editor = {Denise Manahan-Vaughan},
series = {Handbook of Behavioral Neuroscience},
publisher = {Elsevier},
volume = {28},
pages = {509-525},
year = {2018},
booktitle = {Handbook of in Vivo Neural Plasticity Techniques},
issn = {1569-7339},
doi = {https://doi.org/10.1016/B978-0-12-812028-6.00027-6},
url = {https://www.sciencedirect.com/science/article/pii/B9780128120286000276},
author = {Halle R. Dimsdale-Zucker and Charan Ranganath}
}

@article{doi:10.1126/science.1108905,
author = {Tom J. Wills  and Colin Lever  and Francesca Cacucci  and Neil Burgess  and John O'Keefe },
title = {Attractor Dynamics in the Hippocampal Representation of the Local Environment},
journal = {Science},
volume = {308},
number = {5723},
pages = {873-876},
year = {2005},
doi = {10.1126/science.1108905},
URL = {https://www.science.org/doi/abs/10.1126/science.1108905},
eprint = {https://www.science.org/doi/pdf/10.1126/science.1108905}}

@incollection{Kiefer2019-KIERIT,
	pages = {384--409},
	title = {Representation in the Prediction Error Minimization Framework},
	editor = {Sarah K. Robins and John Symons and Paco Calvo},
	booktitle = {The Routledge Companion to Philosophy of Psychology: 2nd Edition},
	year = {2019},
	author = {Alex Kiefer and Jakob Hohwy}
}

\newpage
\section*{Appendix A: Effect of anchor set size on reconstruction}

\begin{figure}[h]
    \centering
    \includegraphics[width=\linewidth]{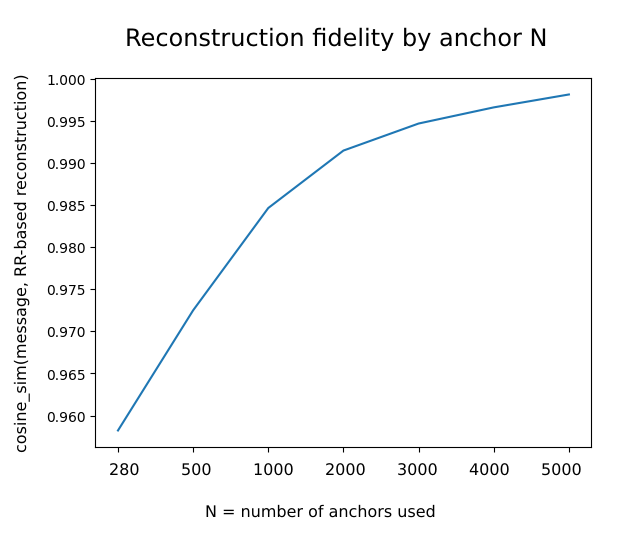}
    \caption{Average cosine similarity ($\frac{u \cdot v}{\lVert u \rVert \lVert v \rVert}$) between ground-truth CSCG beliefs (messages) and their reconstructions from those of a distinct CSCG model trained on the same room and receiving the same sequence of observations, using the method in Equation \ref{reconstruction}, plotted against number $N$ of anchors used to define the relative representations. We begin by setting $N$ to the dimensionality of the model's hidden state. The average is across all 5000 messages in a test sequence.}
    \label{cosine_sim}
\end{figure}

\section*{Appendix B: Visualizing the correspondence of relative representations across models}

\section*{Appendix C: Comparison to LLC}

Locally Linear Coordination (LLC) \cite{NIPS2002_3a1dd983} is a method for aligning the embeddings of multiple dimensionality-reducing models so that they project to the same global coordinate system. While its aims differ somewhat from the procedure outlined in the present study, LLC is also an approach to translating multiple source embeddings to a common representational format. As noted above, there is an interesting formal resemblance between the two approaches, which we explore in this Appendix.

\subsection*{The LLC representation}

LLC presupposes a mixture model of experts trained on $N$ $D$-dimensional input datapoints $\mathcal{X} = [\mathit{x}_1, \mathit{x}_2, ..., \mathit{x}_N]$, in which each expert $m_k$ is a dimensionality reducer that produces a local embedding $z_{n_k} \in \mathbb{R}^{d_k}$ of datapoint $x_n$. The mixture weights or ``responsibilities'' for the model can be derived, for example, as posteriors over each expert's having generated the data, in a probabilistic setting. 

Given the local embeddings and responsibilities, LLC proposes an algorithm for discovering linear mappings $L_k \in \mathbb{R}^{d \times d_k}$ from each expert's embedding to a common (lower-dimensional) output representation $\mathcal{Y} \in \mathbb{R}^{N \times d}$, which can then be expressed as a responsibility-weighted mixture of these projections. That is to say, leaving out bias terms for simplicity: each output image $\mathit{y}_n$ of datapoint $\mathit{x}_n$ is computed as

\begin{equation}
    \mathit{y}_n = \sum_k{r_{n_k}}\big({L_k}{z_{n_k}}\big)
\label{LLC}
\end{equation}

Crucially for what follows, with the help of a flattened (1D) index that spans the ``batch'' dimension $N$ as well as the experts $k$, we can express this in simpler terms as $\mathcal{Y} = UL$. We define matrices $U \in \mathbb{R}^{N \times \sum_k{d_k}}$ and $L \in \mathbb{R}^{\sum_k{d_k} \times d}$ in terms of, respectively: (a) vectors $u_n$, where $u_{n_j} = r_{n_k}z^i_{n_k}$ (i.e. the $j$th element of $u_n$ is the $i$th element of $k$'s embedding of $\mathit{x}_n$ scaled by its responsibility term) --- and (b) re-indexed, transposed columns $l_j = l^i_k$ of the $L_k$ matrices. Intuitively, each row $u_n$ of $U$ concatenates the experts' responsibility-weighted embeddings $r_{n_k}z_{n_k}$ of datapoint $\mathit{x}_n$, while each of $L$'s $d$ columns is a concatenation of the corresponding row of the projection matrices $L_k$, so that the matrix product $UL$ returns a responsibility-weighted prediction for $\mathit{y}_n$ in each row (see Figure \ref{LLC_diagram}).

\subsection*{Relationship to our proposal}

\begin{figure}[h]
    \centering
    \includegraphics[width=\linewidth]{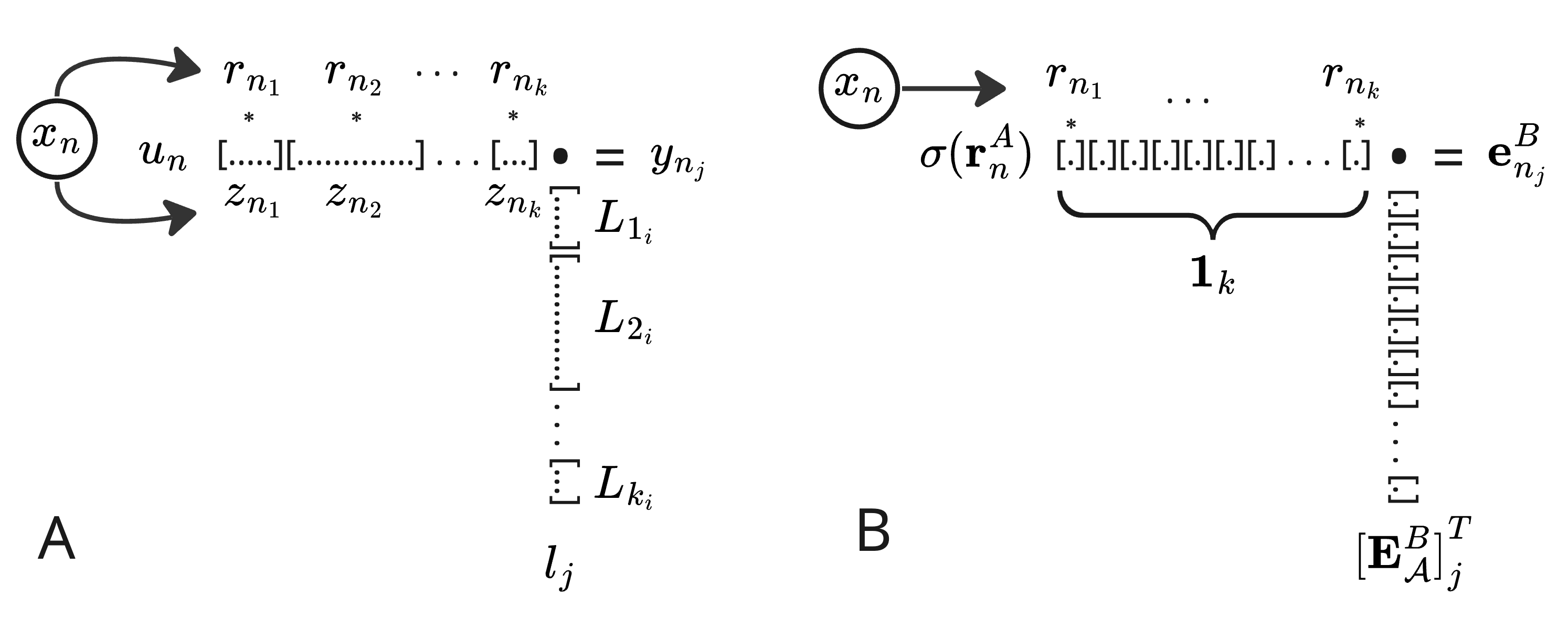}
    \caption{Visual schematic of the computation of a single entry of the output of (A) the projection of input $\mathit{x}_n$ to output $\mathit{y}_n$ as in the Locally Linear Coordinates (LLC) mapping procedure; (B) the reconstruction of a latent embedding $\mathbf{e}^B_n$ in model B's embedding space given input $\mathit{x}_n$ to model A. The groupings in brackets in (A) illustrate the concatenations of vector embeddings (scaled by responsibility terms $r_{n_k}$) in $u_n$, and of projection columns in $l_j$. $\mathbf{1}_k$ in (B) denotes a row of $k$ $1$s (where $k$ in this case denotes the number of anchors, i.e. is set to $\lvert \mathcal{A} \rvert$). Each entry in the column vector ${\big[\mathbf{E}^B_\mathcal{A}\big]}^T_j$ is the $j$th dimension of one of model B's anchor embeddings.}
    \label{LLC_diagram}
    \vspace{-0.5cm}
\end{figure}

Ignoring the motivation of dimensionality reduction which is irrelevant for present purposes, there is a precise conceptual and formal equivalence between this model and the procedure for reconstructing model B's embeddings given those of model A described above in Section 5.2.

Specifically, we can regard each of model A's anchor embeddings $\mathbf{e}^A_{x_k}$ as an "expert" in a fictitious mixture model, with an associated responsibility term measuring its fidelity to the input $\mathit{x_i}$, which in this case is given by the cosine similarity between the anchor embedding and the input embedding. Then like the rows of $U$, each row of $\sigma\big[\mathbf{R}^A_X\big]$, which is a relative representation $\mathbf{r}^A_i = \mathbf{E}^A_\mathcal{A}{\mathbf{e}^A_i}$ of input $i$ after application of the softmax, acts as a responsibility-weighted mixture of multiple ``views'' of the input. Similarly, since the rows of $\mathbf{E}^B_\mathcal{A}$ are anchor embeddings in the output space, its columns $j$ act precisely as do the columns of $L$, i.e. as columns in a projection matrix, so that $\sigma[\mathbf{r}^A_i] \cdot {\mathbf{E}^B_\mathcal{A}}_j$ outputs dimension $j$ of the reconstructed target embedding $\mathbf{e}^B_i$.

There is at least one important difference between LLC and our procedure: in LLC each expert uses an internal transform to generate an input-dependent embedding, which is then scaled by its responsibility term, which also depends on the input. Reconstruction via relative representations instead employs fixed stored embeddings, so that each ``expert'' contributes a scalar value rather than an embedding vector to the final output. However, the expression of LLC in terms of a linear index demonstrates that this makes no essential difference mathematically (conceptually, these scalar ``votes'' are 1D vectors; cf. Figure \ref{LLC_diagram}).

The point is not that these two algorithms are doing precisely the same thing (they are not, as LLC aims to align multiple embedding spaces by deriving a mapping to a distinct common space, while our approach aims to recover the contents of one embedding space from another). The use of LLC to reconstruct input data $\mathcal{X}$ from its ``global'' embedding $\mathcal{Y}$ as in \cite{NIPS2002_3a1dd983} \emph{is} quite closely related to our procedure, however, and at this level of abstraction the approaches may be regarded as the same, with a difference in the nature of the ``experts'' used in the mixture model and the attendant multiple ``views'' of the data. The relative representation reconstruction procedure, while presumably not as expressive, may compensate to some extent for the use of scalar ``embeddings'' by using a large number of ``experts'', and has the virtue of eschewing the need for a mixture model to assign responsibilities, or indeed for multiple intermediate embedding models, to perform such a mapping. 

\end{document}